\documentclass[journal,12pt,onecolumn,a4paper,draftclsnofoot]{IEEEtran}

\usepackage[utf8]{inputenc}
\usepackage[english]{babel}
\usepackage{graphicx}
\usepackage{psfrag}
\usepackage{amsmath}
\usepackage[english,algoruled,commentsnumbered,onelanguage,vlined]{algorithm2e}

\begin{document}

\title{SS5G: Collision Resolution Protocol for Delay and Energy Efficient LoRa Networks}

\author{
\IEEEauthorblockN{Nancy El Rachkidy$^{(1)}$, Alexandre Guitton$^{(1)}$, Megumi Kaneko$^{(2)}$}

\IEEEauthorblockA{
(1) Université Clermont Auvergne, CNRS, LIMOS, F-63000 Clermont-Ferrand, France\\
(2) National Institute of Informatics, Hitotsubashi, 2-1-2, Chiyoda-ku, 101-8430 Tokyo, Japan\\
Emails: nancy.el\_rachkidy@uca.fr, alexandre.guitton@uca.fr, megkaneko@nii.ac.jp}
}

\maketitle

\begin{abstract}
Future 5G and Internet of Things (IoT) applications will heavily rely on long-range communication technologies such as low-power wireless area networks (LPWANs). In particular, LoRaWAN built on LoRa physical layer is gathering increasing interests, both from academia and industries, for enabling low-cost energy efficient IoT wireless sensor networks  for, e.g., environmental monitoring over wide areas. While its communication range may go up to 20 kilometers, the achievable bit rates in LoRaWAN are limited to a few kilobits per second. In the event of collisions, the perceived rate is further reduced due to packet loss and retransmissions. 
Firstly, to alleviate the harmful impacts of collisions, we propose a decoding algorithm that enables to resolve several superposed LoRa signals. Our proposed method exploits the slight desynchronization of superposed signals and specific features of LoRa physical layer. Secondly, we design a full MAC protocol enabling collision resolution. The simulation results demonstrate that the proposed method outperforms conventional LoRaWAN jointly in terms of system throughput, energy efficiency as well as delay. These results show that our scheme is well suited for 5G and IoT systems, as one of their major goals is to provide the best trade-off among these performance objectives.
\end{abstract}


\section{Introduction}
\label{section:introduction}

Long-range low-power communication technologies such as LoRa~\cite{lora}, Sigfox~\cite{sigfox}, and Ingenu~\cite{ingenu}, are becoming widely used in Low-Power Wide Area Networks (LPWANs)~\cite{Xiong15low,Raza17low,Karkatis16demyst}. These technologies enable to cover extensive  zones with very low energy consumption and are thus attractive technologies for supporting the future Internet of Things (IoT) communications and applications, in particular environmental monitoring~\cite{centenaro16long,nolan16evaluation,petajajarvi17evaluation}.

LoRa~\cite{lora} is a recent physical layer for LPWANs making use of Chirp Spread Spectrum (CSS) modulations, which can adaptively extend the communication range by reducing the achievable throughput. On top of this LoRa physical layer, LoRaWAN~\cite{lorawan-2017} defines a simple MAC protocol based on open specification, which allows end-devices to communicate to a network server through gateways, but with a small duty-cycle (e.g., 1\%). Thus, end-devices can save energy, and the network lifetime is increased.
The main issue in LoRa and LoRaWAN is their throughput limitation: the indicative physical bitrate varies between 250 and 11000 bps~\cite{lorawan-regional-settings-2017}. Moreover, when two end-devices transmit simultaneously using the same parameters such as channel, Spreading Factor (SF), and are received by the gateway with a similar power, a collision occurs and none of the signals are decoded by LoRa. Thus, both end-devices have to retransmit, which further reduces their achievable throughput.

So far, a number of works have focused on channel and SF allocation issues for the uplink transmissions of LoRa systems, among which~\cite{Qin17spreading, Reynders17power, Lim18spreading}. Most of these methods rely on a centralized scheduling unit at the gateway. The feasibility of large-scale LoRa networks has been analyzed in~\cite{Bor16nov, Geo17apr}, in particular the effect of co-SF interferences as a large number of end-devices may use the same SF at the same time. Most previous works consider SFs to be orthogonal, but recently, various experiments and analysis have pointed out the impact of imperfect orthogonality of SFs whereby devices using different SFs may interfere among themselves~\cite{croce17impact, zhu18evaluation, Waret18LoRa}. 

To alleviate the large performance degradations due to co-SF interferences, we have proposed in~\cite{elrachkidy18decoding} a method for decoding superposed LoRa signals by exploiting the specific features of LoRa signals. The proposed algorithm was shown to provide significant performance enhancements in terms of achievable throughput, for different SF levels. However, the algorithm in~\cite{elrachkidy18decoding} was solely designed to handle two superposed LoRa signals and we did not consider any MAC protocol. 

Therefore, in this work, we extend our preliminary proposal of~\cite{elrachkidy18decoding} by designing a general decoding algorithm for several signals, which is far more intricate than the restrictive case of two superposed signals. In addition, we propose a tailored MAC protocol on top of our decoding algorithm.
In particular, we show that it is possible to retrieve the frames from superposed signals that are slightly desynchronized, with reasonable assumptions on the hardware.

Our contributions are three-fold. Firstly, we propose an algorithm that is able to cancel the collision between two collided signals and thus retrieve entire frames without any loss. We then generalize this algorithm for retrieving several collided frames that are sent by several end-devices. Secondly, we propose a MAC layer slotted with beacons in order to allow synchronized transmissions (and to compensate for the drifting of the end-devices). This MAC layer divides time into slots in which several end-devices might send slightly desynchronized frames. Thirdly, we propose a Cyclic Redundancy Check (CRC) decoding scheme that can be applied in order to decode the few frames that our algorithm was unable to decode.

The structure of this paper is as follows. Section~\ref{section:state-of-the-art} describes the LoRaWAN technology with the LoRa physical layer and the LoRaWAN MAC layer. Section~\ref{section:proposition} presents the proposed decoding algorithm designed to correctly decode the collided frames, followed by the proposed MAC layer in Section~\ref{section:new-mac}. Section~\ref{section:results} shows the simulation parameters we use and the results we obtained. Finally, Section~\ref{section:conclusion} concludes the paper.

\section{LoRaWAN Description}
\label{section:state-of-the-art}

In the following, we first describe the LoRa physical layer, which is the main focus of our paper. Then, we describe the LoRaWAN MAC protocol.

\subsection{LoRa}
\label{subsection:lora}

LoRa~\cite{lora} is a physical layer technology for LPWAN, based on a Chirp-Spread Spectrum (CSS) modulation. Each LoRa chirp consists of a linear frequency sweep. The duration of the sweep is called symbol duration (SD), and depends on the value of the spreading factor $SF$ and on the bandwidth $BW$. The sweep is performed over the whole bandwidth $BW$. Chirps are either up-chirps, where the frequency sweep is increasing, and down-chirps, where the frequency sweep is decreasing.

\begin{figure}[htbp]
    \centering
    \psfrag{frequency}[c]{{\small frequency}}
    \psfrag{BW}[c]{{\small $BW$}}
    \psfrag{value}[c]{{\small value}}
    \psfrag{SD}[c]{{\small $SD$}}
    \psfrag{time}[c]{{\small time}}
    \includegraphics[scale=1.4]{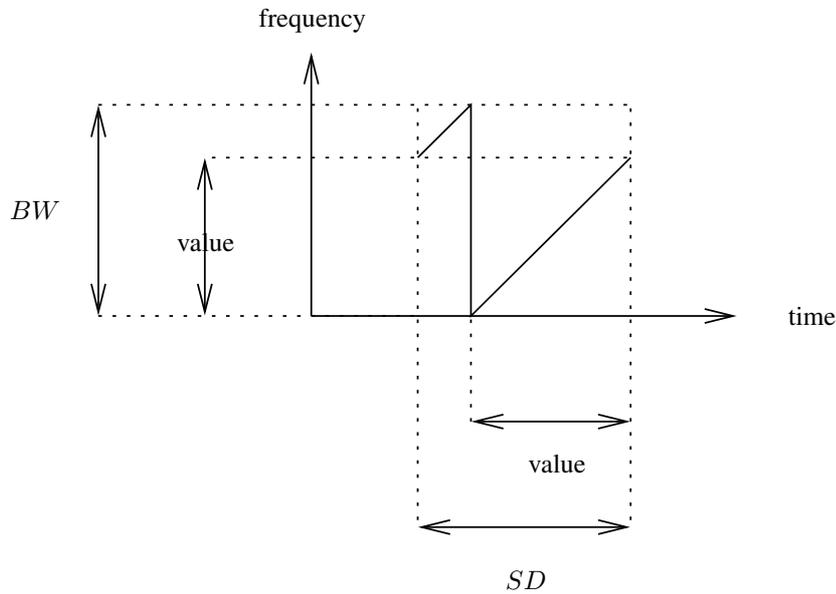}
    \caption{Example of a single LoRa up-chirp. Computing the symbol value requires knowing the symbol start time and the initial frequency, or the sharp frequency edge and the symbol end time.}
    \label{figure:one-symbol}
\end{figure}

Each chirp is a symbol and can encode $2^{SF}$ possible values. This is achieved by shifting the sweep by the symbol value, as shown on Figure~\ref{figure:one-symbol} for an up-chirp. From the sharp edge in the instantaneous frequency trajectory~\cite{goursaud15dedicated}, and assuming that the receiver is synchronized with the transmitter, the receiver can compute the symbol value as the shift in the frequency at the beginning of the symbol. The symbol value of an up-chirp is also proportional to the remaining time between the sharp frequency edge and the end of the symbol, as shown on Figure~\ref{figure:one-symbol}. The symbol value of a down-chirp is proportional to the time between the beginning of the symbol and the sharp frequency edge.

To decode a symbol, the receiver needs to know the frontier of the symbol. Thus, LoRa synchronizes the transmitter and the receiver by using a preamble of a few symbols. In the case of uplink communications, the preamble consists of three parts: (i) a series of up-chirps (generally six), each having a symbol value of 0, (ii) two up-chirps encoding the sync word, which is a network identification, and (iii) two and a quarter down-chirps, used to identify the end of the preamble. The payload and a CRC follow the preamble, and are encoded using up-chirps. LoRa allows an explicit header mode, which inserts a header between the preamble and the payload. This header contains the payload length, the coding rate, and an optional header CRC.

\begin{figure}[htbp]
    \centering
    \psfrag{sender}[c]{{\small sender}}
    \psfrag{receiver}[c]{{\small receiver}}
    \psfrag{preamble}[c]{{\small preamble}}
    \psfrag{s1}[c]{{\small 6}}
    \psfrag{s2}[c]{{\small 0}}
    \psfrag{s3}[c]{{\small 4}}
    \psfrag{s4}[c]{{\small 4}}
    \psfrag{desynchronization information}[c]{{\small desynchronization information}}
	\includegraphics[scale=0.8]{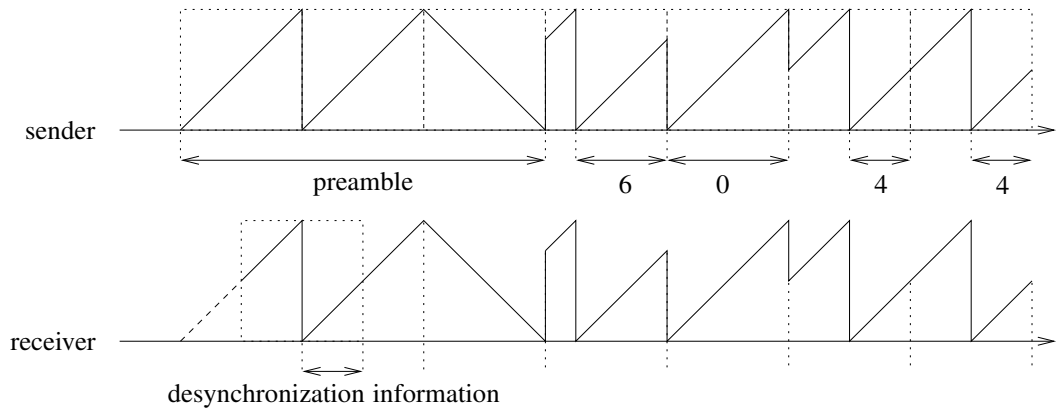}
	\caption{Example of a LoRa uplink frame, with a short preamble and four data symbols, with SF3. The receiver synchronizes itself with the sender during the preamble.}
	\label{figure:one-signal}
\end{figure}

Figure~\ref{figure:one-signal} shows an example of an uplink communication with a shorten preamble (two up-chirps instead of six, no sync word, and one down-chirp instead of two and a quarter) and a few data symbols (four symbols). We chose SF3 for the sake of simplicity, leading to $2^{SF}=8$ possible values per symbol. Let us assume that a desynchronized node starts receiving the preamble, not necessarily at the exact beginning of the preamble. The node detects a sharp frequency edge of the preamble, which indicates the frontier of a symbol. From this information, the receiver can synchronize itself according to the transmitter. The end of the preamble is detected by the inversion of the chirps. Then, the payload is decoded. In this example, the data symbols are 6, 0, 4, 4.

\subsection{LoRaWAN}
\label{subsection:lorawan}

LoRaWAN (in version 1.0~\cite{lorawan-2015} or in version 1.1~\cite{lorawan-2017}) is a simple MAC layer. It is based on the LoRa physical layer. The topology defined in LoRaWAN is a star topology where end-devices are connected to a network server through relays called gateways. The communication technology between the end-devices and the gateways is based on CSS modulations. Moreover, LoRaWAN defines three classes for end-devices: class A is for low-power uplink communications, class B is for delay-guaranteed downlink communications, and class C is for end-devices without energy constraints. In class A, which is the only mandatory class, the end-devices are energy-efficient. In this class, the end-devices can transmit at any time using ALOHA mechanism: an end-device chooses a channel randomly, sends the frame, and waits for an acknowledgement during two successive receive windows. The transmission time of each end-device should not exceed 1\%.

LoRaWAN manages the bitrate according to the quality of links. Indeed, it uses the SF of the signal in order to have a trade-off between the robustness of the signal and the bitrate. When an end-device experiences a low signal quality, it increases its SF in order to be able to send frames over long distances and thus better decode the signal. However, this results into lower bitrate. This adaptation is controlled by the datarate (DR) of LoRaWAN, which varies from DR0 (for large SF but small bitrate) to DR6 (for small SF but larger bitrate).

The European regional settings of LoRaWAN~\cite{lorawan-regional-settings-2017} define most LoRa parameters. The bandwidth of channels, $BW$, is equal to 125 kHz for DR0 to DR5, and 250 kHz for DR6. SF varies from 12 down to 7 for DR0 to DR5, and is equal to 7 for DR6. The preamble length is equal to 6 symbols. The physical bitrate varies between 250 bps for DR0, to 11000 bps for DR6. The maximum MAC payload of a frame varies between 59 bytes for DR0 and 230 bytes for DR6.

\section{Proposed Superposed LoRa Signal Decoding}
\label{section:proposition}


LoRa gateways are able to decode superposed LoRa signals as long as they are sent on different channels or on different SFs. Notice however that some researchers have shown that signals on different SFs are not completely orthogonal~\cite{croce17impact,zhu18evaluation,Waret18LoRa}.

When several signals are received on the same channel and with the same SF, a difference of received power might cause the strongest signal to be captured~\cite{goursaud15dedicated,haxhibeqiri17lora}. When several signals have a similar receive power, a collision occurs and all signals are considered lost~\cite{Bor16nov,Geo17apr}.

In this paper, we focus on decoding superposed LoRa signals of {\it similar receive power}, on the {\it same channel}, with the {\it same SF}. To do so, we show that we can use timing information to match the correct symbols to the correct end-device.

In Subsection~\ref{subsection:assumptions}, we describe our assumptions. In Subsection~\ref{subsection:two-signals}, we provide our main algorithm, and we describe how it can decode two signals that are slightly desynchronized. In Subsection~\ref{subsection:three-signals}, we extend the algorithm for the case of three or more signals that are slightly desynchronized. In Subsection~\ref{subsection:crc}, we show how the CRC of frames can be used to decode additional frames. 

\subsection{Assumptions}
\label{subsection:assumptions}

As in~\cite{elrachkidy18decoding}, we assume that there are no non-linearity effects between up-chirps (respectively down-chirps). In other words, if two up-chirps (resp. down-chirps) $c_1$ and $c_2$ overlap at a given time $t$ at the receiver side, the two observed frequencies are the frequency of $c_1$ (at time $t$) and the frequency of $c_2$ (at time $t$). Without additional information, it is not possible to correlate the frequency to the corresponding transmitter. We assume that when an up-chirp is superposed with a down-chirp, it is not possible to detect any of the frequencies. We assume that when several frequencies overlap at a given time, only one frequency is detected by the receiver. For instance, if there are three nodes transmitting at a given time, but only two frequencies $f_1$ and $f_2$ are detected, we assume that it is not possible to know whether two nodes were transmitting with $f_1$ and one with $f_2$, or one node was transmitting with $f_1$ and two with $f_2$.

We assume that it is possible for the hardware of the receiver to detect all frequencies of overlapping up-chirps (resp. down-chirps) within $\delta$ time-units. In the following examples, we use $\delta=SD/4$ unless stated otherwise. Please note that on real LoRa hardware, the decoding of signals is not carried out by directly detecting the sharp frequency edges, but instead by computing a fast Fourier transform and detecting the peak in the frequency domain~\cite{goursaud15dedicated}. With our proposition, this translates into either detecting the two sharp frequency edges in the time domain, or the two peaks in the frequency domain. In practice, it is likely that $\delta$ cannot be too small, as uncertainties in frequency detection might occur.

We also assume some properties on the frames: all nodes transmit with the same preamble duration, the frame length is included in the explicit header, and there is at least one symbol change during the whole frame: that is, the payload (data and CRC) does not consist of a sequence of identical symbols.

Finally, we consider that nodes are slightly desynchronized: all nodes start their transmission within $SD-\delta$ time units, and during the whole transmission duration, the transmissions of any two nodes have a delay of at least $\delta$ time units. In the following examples, we assume that each node $n_i$ starts transmitting at time $t_0+(i-1)\delta$ (for $i\geq{}1$), and we consider that time drift between transmitters is negligible as the time on air of LoRa frames is short.

\subsection{Case of two slightly desynchronized signals}
\label{subsection:two-signals}

In this subsection, we consider the superposition of two signals from two transmitters that are slightly desynchronized (by at least $\delta$ time units, and at most $SD-\delta$ time units).

\begin{figure*}[htbp]
    \centering
    \psfrag{t0}[c]{$t_0$}
    \psfrag{t1}[c]{$t_1$}
    \psfrag{t2}[c]{$t_2$}
    \psfrag{t3}[c]{$t_3$}
    \psfrag{t4}[c]{$t_4$}
    \psfrag{t5}[c]{$t_5$}
    \psfrag{t6}[c]{$t_6$}
    \psfrag{t7}[c]{$t_7$}
    \psfrag{t8}[c]{$t_8$}
    \psfrag{t9}[c]{$t_9$}
    \psfrag{t10}[c]{$t_{10}$}
    \psfrag{t11}[c]{$t_{11}$}
    \psfrag{t12}[c]{$t_{12}$}
    \psfrag{n1}[c]{$n_1$}
    \psfrag{n2}[c]{$n_2$}
    \psfrag{0}[c]{{\small 0}}
    \psfrag{1}[c]{{\small 2}}
    \psfrag{2}[c]{{\small 4}}
    \psfrag{3}[c]{{\small 6}}
    \psfrag{receiver}[c]{receiver}
    \psfrag{delta}[c]{$\delta$}
    \includegraphics[scale=0.8]{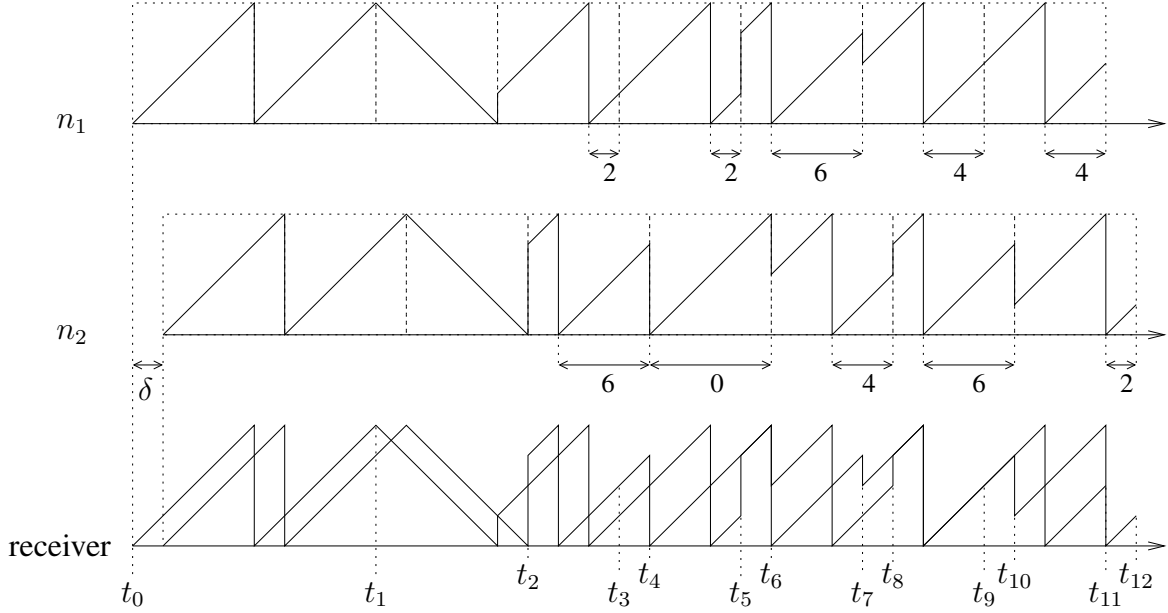}
    \caption{The superposition of two slightly desynchronized signals produces a complex signal, which can still be decoded in linear time.}
    \label{figure:two-signals}
\end{figure*}

Figure~\ref{figure:two-signals} shows the superposition of two slightly desynchronized signals. The preamble length is three symbols (2 up-chirps instead of 6, no sync word, and 1 down-chirp instead of 2.25), and SF3. The figure shows the signal of the first transmitter $n_1$ starting at $t_0$, the signal of the second transmitter $n_2$ starting at $t_0+\delta$, and the superposed signal at the receiver. Note that the data transmitted by $n_1$ is $(2,2,6,4,4)$, and the data transmitted by $n_2$ is $(6,0,4,6,2)$. We will first explain our algorithm on this example, and then proceed with a more formal description.

\begin{table*}[htbp]
    \centering
    \begin{tabular}{|c|c|c|c||c|c|c|c|}
        \hline
        time & $F_-$ & $F_+$ & symbol & time & $F_-$ & $F_+$ & symbol\\
        \hline
        $t_2$ & unknown & $\{4,6\}$ & initialization
        & $t_3$ & $\{2,4\}$ & $\{2,4\}$ & $s^1_1=*,s^1_2=s^1_1$\\
        $t_4$ & $\{4,6\}$ & $\{0,4\}$ & $s^2_1=6,s^2_2=0$ 
        & $t_5$ & $\{2,6\}$ & $\{6\}$ & $s^1_2=2,s^1_3=6$\\
         $t_6$ & $\{0\}$ & $\{0,4\}$ & $s^2_2=0,s^2_3=4$
        & $t_7$ & $\{2,6\}$ & $\{2,4\}$ & $s^1_3=6,s^1_4=4$ \\
        $t_8$ & $\{4,6\}$ & $\{6\}$ & $s^2_3=4,s^2_4=6$
        & $t_9$ & $\{4\}$ & $\{4\}$ & $s^1_5=s^1_4$ \\
         $t_{10}$ & $\{6\}$ & $\{2,6\}$ & $s^2_4=6,s^2_5=2$
        &$t_{11}$ & $\{0,4\}$ & $\{0\}$ & $s^1_5=4,s^1_6=0$\\
         $t_{12}$ & $\{2\}$ & $\emptyset$ & $s^2_5=2$
        & & & & \\
        \hline
    \end{tabular}
    \caption{Decoding of the two signals of Figure~\ref{figure:two-signals}.}
    \label{table:two-signals-explanation}
\end{table*}

\underline{\it Example of preamble detection and data decoding}

{\it Preamble detection:} During $[t_0;t_0+\delta]$, the receiver detects the preamble of $n_1$. During $[t_0+\delta;t_0+2\delta]$, the receiver is able to detect that two slightly desynchronized signals are transmitted, and is able to deduce the symbol frontiers of both transmitters. At frontier $t_1$, or more precisely, during $[t_1;t_1+\delta]$, the receiver is not able to detect the superposition of preambles anymore (due to the presence of up-chirps superposed with down-chirps). Thus, it knows that the preamble of $n_1$ has reached its first down-chirp at $t_1$.

{\it Data decoding:} We define the sequence of decoded data for $n_1$ by $s^1$ and the sequence of decoded data for $n_2$ by $s^2$. $t_2$, which is the beginning of the payload of $n_2$, is the first time where only up-chirps of data symbols are superposed. At frontier $t_2$, the receiver stores the current frequencies, which correspond to $F_{+}(t_2)=\{4,6\}$. At frontier $t_3$, the receiver computes $F_{-}(t_3)$ by updating the previous frequencies $F_{+}(t_2)=\{4,6\}$, and obtains $F_{-}(t_3)=\{2,4\}$ (each frequency of $F_{+}(t_2)$ is increased by $3/4\cdot{}2^{SF}=6$ since $3/4$ time units have passed since $t_2$). The receiver detects the current frequencies $F_{+}(t_3)=\{2,4\}$. There is no change in the frequencies ($F_{-}(t_3)=F_{+}(t_3)$), since the beginning of the data of $n_1$ starts with the repeated symbol 2. Thus, the algorithm leaves $*$ for the first symbol of $n_1$ (to be decoded later), so $s^1=(*)$. At frontier $t_4$, the receiver computes $F_{-}(t_4)$ by updating the previous frequencies $F_{+}(t_3)=\{2,4\}$, and obtains $F_{-}(t_4)=\{4,6\}$ (since $1/4$ time units have passed). It detects the current frequencies $F_{+}(t_4)$, and obtains $F_{+}(t_4)=\{0,4\}$. Thus, one frequency changed from 6 to 0, hence, $s^2=(6,0)$, since $t_4$ is a frontier of $n_2$.
The current symbol of $n_1$ corresponds to frequency 4 (which is translated into 2 at the beginning of the symbol frontier of $n_1$, which was $t_3$). At frontier $t_5$, the receiver computes $F_{-}(t_5)$ by updating the previous frequencies $F_{+}(t_4)=\{0,4\}$, and obtains $F_{-}(t_5)=\{2,6\}$. It detects the current frequencies $F_{+}(t_5)=\{6\}$, which can also be written $\{6,6\}$. The frequency of $n_1$ changed from 2 to 6, hence $s^1=(*,2,6)$. The current symbol of $n_2$ corresponds to frequency 6 (which translates to 0 at the beginning of the symbol frontier of $n_2$, $t_4$, and was already known). The algorithm continues until $t_{12}$, where no frequency is received. Thus, the algorithm knows that all nodes have stopped their transmissions. The algorithm removes the last predicted symbol of $n_1$ (indeed, at $t_{11}$, it considered that $n_1$ was transmitting a symbol with the same frequency as the frequency of $n_2$). At this step, the decoded frames are $s^1=(*,2,6,4,4)$ for $n_1$ and $s^2=(6,0,4,6,2)$ for $n_2$. Then, the algorithm replaces all special values $*$ with the first known value of the frame by backtracking (since we know $s^1_2=s^1_1$). The algorithm uses the frame length present in each frame to truncate the frames to their correct length. Finally, the algorithm outputs are $(2,2,6,4,4)$ and $(6,0,4,6,2)$, as expected.

%

\underline{\it Generalization of preamble detection and data decoding}

In this paragraph, we generalize the example given above and we formulate our proposition in Algorithm~\ref{algorithm:proposition}. 

{\it Preamble detection:} The superposition of the beginning of the preambles results in the superposition of up-chirp symbols. This superposition enables the receiver to detect two sharp frequency edges, each sharp edge allowing the receiver to know the symbol frontier of a transmitter. The beginning of the first data symbol of the first node is not decodable, as it corresponds to an up-chirp (for node $n_1$) superposed with a down-chirp (for the end of the preamble of $n_2$).

{\it Data decoding:} From the beginning of the first data symbol of the second node, only up-chirps are superposed, and thus it is possible to detect all sharp edges. The difficulty relies in correlating each frequency with the symbols of each node. To do so, we use the following property: sharp edges can occur only at the beginning of a symbol, when the symbol changes, or once during a symbol. When the sharp edge occurs during a symbol, it can be predicted if the symbol value is known.

Algorithm~\ref{algorithm:proposition} describes our proposed algorithm. It starts after the superposed preambles have been received, and thus considers that the symbol frontier of each transmitter is known. The algorithm considers the frontiers of all data symbols sequentially, apart from the first frontier of the first node for which the frequency cannot be obtained. At each frontier, the receiver updates the previous frequencies (since frequencies change over time in LoRa chirps, and time has passed since the detection of the previous frequencies). Then, the receiver compares these (updated) previous frequencies $F_-$ with the current frequencies $F_+$. Note that in practice, it may take up to $\delta$ time units to obtain the current frequencies, so the receiver might have to update the current frequencies based on the detection duration. Only two cases can occur for the algorithm.\\
{\it Case 1:} Exactly one frequency has changed. This can only happen when a new symbol starts, which can only occur at the symbol frontier. Since the receiver knows if the current frontier is for the first or the second transmitter, it knows the new symbol for the current node (based on the new frequency), the previous symbol for the current node (based on the frequency that has changed), and the current symbol for the other node (based on the frequency that did not change).\\
{\it Case 2:} No frequency has changed. This can only happen when the new symbol is equal to the previous symbol (this was the case on Figure~\ref{figure:two-signals} at times $t_3$ and $t_9$).
\begin{itemize}
        \item{}If the receiver knows the previous symbol of the current node (time $t_9$ of Figure~\ref{figure:two-signals}), the new symbol can be deduced.
        \item{}Otherwise, the previous symbol of the current node is unknown, which corresponds to the beginning of the algorithm when the first symbol is repeated (time $t_3$ of Figure~\ref{figure:two-signals}). In this case, the algorithm leaves a special value (denoted by $*$ here). As soon as one symbol changes, the receiver is able to identify the new and previous symbols of the end-device corresponding to that frontier, and hence to deduce the symbol of the other end-device. In addition, the algorithm can replace all the $*$ values of the frame of the current node with the value of the previous symbol. This is why we assumed at least one symbol change per frame.
\end{itemize}

\begin{algorithm}[htbp]
    \For{each frontier $t_i$ of a data chirp}{
        compute currentSymbol and currentNode\\
        \If{currentSymbol=0 and currentNode=1}{
            skip (frequencies cannot be detected)\\
        }
        \Else{
            $F_{+}(t_i)\leftarrow$detect current frequencies\\
            \If{currentSymbol=0 and currentNode=2}{
                skip ($F_{-}(t_i)$ cannot be computed)\\
            }
            \Else{
                compute $F_{-}(t_{i})$ by updating $F_{+}(t_{i-1})$\\
                $newF\leftarrow{}F_{+}(t_i)-F_{-}(t_i)$\\
                $oldF\leftarrow{}F_{-}(t_i)-F_{+}(t_i)$\\
                \If{$newF=\emptyset$}{
                    the new symbol in $symb[$currentNode$]$ is equal to the previous (or to $*$)
                }
                \Else{
                    the previous symb. in $symb[$currentNode$]$ is equal to the value of $oldF$\\
                    the new symbol in $symb[$currentNode$]$ is equal to the value of $newF$
                }
            }
        }
    }
    \For{each node $n$}{
        replace in $symb[n]$ all the leading $*$ values with the first defined value\\
        truncate the frame according to its length
    }
    \caption{Decoding of two slightly desynchronized superposed LoRa signals.}
    \label{algorithm:proposition}
\end{algorithm}

The time complexity of our algorithm is linear with the number of symbols of the longest frame. Most of the symbols are decoded on the fly, $\delta$ time units after the beginning of the symbol, except for the symbols repeated initially (see the last loop of the algorithm). The space complexity of our algorithm is $\mathcal{O}(1)$, since the storage requirement is limited to the value of the first non-special symbol for each node. Thus, the algorithm is extremely efficient in time and space, for two nodes.

\subsection{Case of several slightly desynchronized signals}
\label{subsection:three-signals}

\begin{figure}[htbp]
	\centering
    \psfrag{n1}[c]{{\small $n_1$}}
    \psfrag{n2}[c]{{\small $n_2$}}
    \psfrag{n3}[c]{{\small $n_3$}}
    \psfrag{receiver}[c]{{\small receiv.}}
    \psfrag{0}[c]{{\small 0}}
    \psfrag{1}[c]{{\small 2}}
    \psfrag{2}[c]{{\small 4}}
    \psfrag{3}[c]{{\small 6}}
	\includegraphics[scale=0.8]{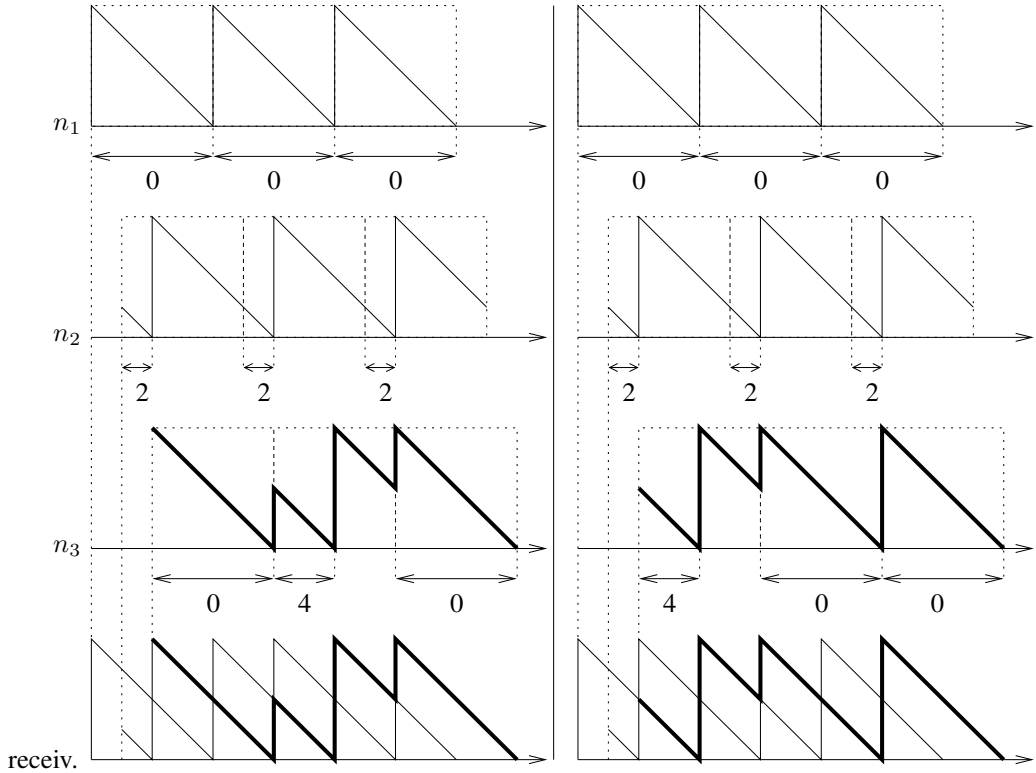}
	\caption{When three nodes that are slightly desynchronized transmit frames, it is not always possible to decode them: these two sets of frames produce the same superposition of frequencies.}
	\label{figure:undecodable}
\end{figure}

Note that with our assumptions, decoding three or more signals is not always possible. For instance, Figure~\ref{figure:undecodable} shows two sets of different signals that produce the same superposition of frequencies, and thus cannot be decoded.

Algorithm~\ref{algorithm:proposition2} describes our proposed algorithm, for three or more nodes. It is similar to Algorithm~\ref{algorithm:proposition}, with the following main changes. (1) When $F_{-}(t)=F_{+}(t)$ at the frontier of a node $n$, it is not possible to assume that the symbol of $n$ remains the same. Indeed, if the number of frequencies of $F_{-}(t)$ is smaller than the number of nodes, the frequency of node $n$ might have changed from one superposed frequency to another superposed frequency. (2) Consequently, initial repeated symbols which yielded unchanging frequencies cannot be decoded.

\begin{algorithm}[htbp]
    \For{each frontier $t_i$ of a data chirp}{
        compute currentSymbol and currentNode\\
        \If{currentSymbol=0 and currentNode$\neq{}$lastNode}{
            skip (frequencies cannot be detected)\\
        }
        \Else{
            $F_{+}(t_i)\leftarrow$detect current frequencies\\
            \If{currentSymbol=0 and currentNode=lastNode}{
                skip ($F_{-}(t_i)$ cannot be computed)\\
            }
            \Else{
                compute $F_{-}(t_{i})$ by updating $F_{+}(t_{i-1})$\\
                $newF\leftarrow{}F_{+}(t_i)-F_{-}(t_i)$\\
                $oldF\leftarrow{}F_{-}(t_i)-F_{+}(t_i)$\\
                \If{$oldF\neq{}\emptyset$}{
                    the previous symb. in $symb[$currentN.$]$ is equal to the value of $oldF$
                }
                \If{$newF\neq\emptyset$}{
                    the new symbol in $symb[$currentNode$]$ is equal to the value of $newF$
                }
            }
        }
    }
    \caption{Decoding of three or more slightly desynchronized superposed LoRa signals.}
    \label{algorithm:proposition2}
\end{algorithm}

Algorithm~\ref{algorithm:proposition2} is able to decode many cases of slightly desynchronized signals for $n$ transmitters, when $n\geq{}3$, while Algorithm~\ref{algorithm:proposition} is able to decode all cases of slightly desynchronized signals for $n=2$ transmitters. It only fails to do so when the number of received frequencies is within $[2;n-1]$ (which never occurs when $n=2$). Indeed, in this case, even if the algorithm knows that the frequency of the current node has changed, it cannot determine what is the new value, as it has $n-1>1$ possibilities. It can still deduce the value of the previous symbol for this node. At the next frontier for this node, though, the value of this symbol might be deduced, depending on the number of other frequencies.

\begin{figure*}[htbp]
    \centering
    \psfrag{t0}[c]{$t_0$}
    \psfrag{t1}[c]{$t_1$}
    \psfrag{t2}[c]{$t_2$}
    \psfrag{t3}[c]{$t_3$}
    \psfrag{t4}[c]{$t_4$}
    \psfrag{t5}[c]{$t_5$}
    \psfrag{t6}[c]{$t_6$}
    \psfrag{t7}[c]{$t_7$}
    \psfrag{t8}[c]{$t_8$}
    \psfrag{t9}[c]{$t_9$}
    \psfrag{t10}[c]{$t_{10}$}
    \psfrag{t11}[c]{$t_{11}$}
    \psfrag{t12}[c]{$t_{12}$}
    \psfrag{t13}[c]{$t_{13}$}
    \psfrag{t14}[c]{$t_{14}$}
    \psfrag{t15}[c]{$t_{15}$}
    \psfrag{t16}[c]{$t_{16}$}
    \psfrag{t17}[c]{$t_{17}$}
    \psfrag{n1}[c]{$n_1$}
    \psfrag{n2}[c]{$n_2$}
    \psfrag{n3}[c]{$n_3$}
    \psfrag{receiver}[c]{receiver}
    \psfrag{delta}[c]{$\delta$}
    \psfrag{0}[c]{{\small 0}}
    \psfrag{1}[c]{{\small 1}}
    \psfrag{2}[c]{{\small 2}}
    \psfrag{3}[c]{{\small 3}}
    \psfrag{4}[c]{{\small 4}}
    \psfrag{5}[c]{{\small 5}}
    \psfrag{6}[c]{{\small 6}}
    \psfrag{7}[c]{{\small 7}}
    \includegraphics[scale=0.7]{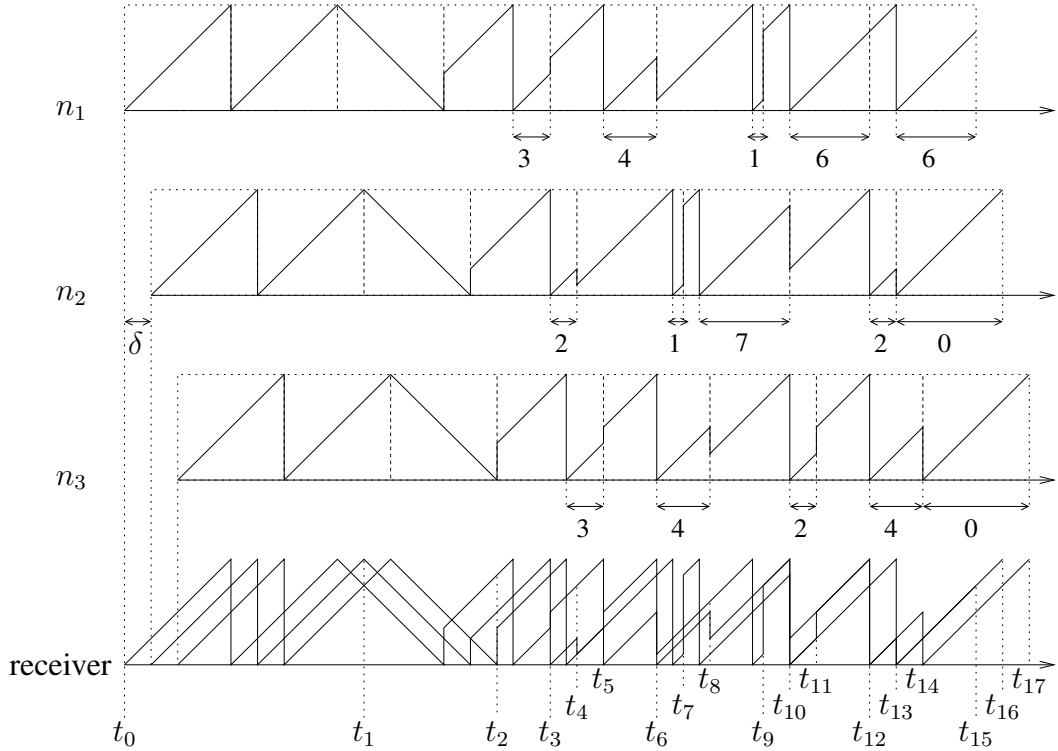}
    \caption{The superposition of three signals produce a very complex signal, which can be partially decoded.}
    \label{figure:three-signals}
\end{figure*}

Figure~\ref{figure:three-signals} shows the superposition of three signals, and Table~\ref{table:three-signals-explanation} shows the decoding of the three superposed signals of Figure~\ref{figure:three-signals}, according to Algorithm~\ref{algorithm:proposition2}. Initially, $F_{+}(t_2)=\{3,4,7\}$. Then, the algorithm computes $F_{-}(t_3)=\{0,3,7\}$ and obtains $F_{+}(t_3)=\{0,4,7\}$. The first symbol $s^1_1$ of node $n_1$ is thus 3, and the second symbol $s^1_2$ of node $n_1$ is 4. Then, the algorithm computes $F_{-}(t_4)=\{1,2,6\}$ and obtains $F_{+}(t_4)=\{1,6\}$. The first symbol $s^2_1$ of node $n_2$ is 2, but it is not possible to determine the second symbol of node $n_2$ yet. Then, the algorithm computes $F_{-}(t_5)=\{0,3\}$ and obtains $F_{+}(t_5)=\{0,3,4\}$. The second symbol $s^3_2$ of node $n_3$ is 4, but it is not possible to determine whether the first symbol of node $n_3$ is 0 or 3. The algorithm continues until $t_{17}$.

\begin{table*}[htbp]
    \centering
    \begin{tabular}{|c|c|c|c||c|c|c|c|}
        \hline
        time & $F_-$ & $F_+$ & symbol & time & $F_-$ & $F_+$ & symbol \\ 
        \hline
        $t_2$ & unknown & $\{3,4,7\}$ & initialization
        & $t_3$ & $\{0,3,7\}$ & $\{0,4,7\}$ & $s^1_1=3,s^1_2=4$ \\
         $t_4$ & $\{1,2,6\}$ & $\{1,6\}$ & $s^2_1=2$ 
        &$t_5$ & $\{0,3\}$ & $\{0,3,4\}$ & $s^3_2=4$ \\
         $t_6$ & $\{0,4,7\}$ & $\{0,1,7\}$ & $s^1_2=4,s^1_3=1$
        & $t_7$ & $\{1,2,3\}$ & $\{2,3,7\}$ & $s^2_2=1,s^2_3=7$ \\
        $t_8$ & $\{1,4,5\}$ & $\{1,2,5\}$ & $s^3_2=4,s^3_3=2$
        & $t_9$ & $\{1,5,6\}$ & $\{5,6\}$ & $s^1_4=1$ \\
         $t_{10}$ & $\{0,7\}$ & $\{0,2\}$ & $s^2_3=7,s^2_4=2$ 
        &$t_{11}$ & $\{2,4\}$ & $\{2,4\}$ & ? \\
         $t_{12}$ & $\{0,6\}$ & $\{0,6\}$ & ?
        & $t_{13}$ & $\{0,2\}$ & $\{0,2\}$ & ? \\
        $t_{14}$ & $\{2,4\}$ & $\{0,2\}$ & $s^3_4=4,s^3_5=0$
        & $t_{15}$ & $\{4,6\}$ & $\{4,6\}$ & ? \\
         $t_{16}$ & $\{0,6\}$ & $\{6\}$ & $s^2_5=0,s^2_6=6$
        &$t_{17}$ & $\{0\}$ & $\emptyset$ & $s^3_5=0$ \\
        \hline
    \end{tabular}
    \caption{Partial decoding of the three signals of Figure~\ref{figure:three-signals}.}
    \label{table:three-signals-explanation}
\end{table*}

Table~\ref{table:three-signals-decoded} shows the output of Algorithm~\ref{algorithm:proposition2}. The frame of $n_2$ is successfully decoded. However, the frame of $n_1$ has its last two symbols unknown, and the frame of $n_3$ has its first symbol unknown.

\begin{table}[htbp]
    \centering
    \begin{tabular}{|c|c|c|c|c|c|}
        \hline
        node & symbol 1 & symbol 2 & symbol 3 & symbol 4 & symbol 5\\
        \hline
        $n_1$ & 3 & 4 & 1 & $\{5,6\}$ & $\{0,6\}$\\ 
        $n_2$ & 2 & 1 & 7 & 2 & 0 \\
        $n_3$ & $\{0,3\}$ & 4 & 2 & 4 & 0\\ 
        \hline
    \end{tabular}
    \caption{Output of Algorithm~\ref{algorithm:proposition2} on the signals of Figure~\ref{figure:three-signals}. Only one frame is completely decoded.}
    \label{table:three-signals-decoded}
\end{table}

\subsection{Cyclic Redundancy Check for decoding}
\label{subsection:crc}

It is possible to use the CRC present in each frame in order to improve the decoding rate of Algorithm~\ref{algorithm:proposition2}.

Let us consider the output of Table~\ref{table:three-signals-decoded} as an example. The first symbol of the frame of $n_3$ is unknown, but the uncertainty is limited to two possible values for this symbol. Thus, the frame for $n_3$ is either (0,4,2,4,0) or (3,4,2,4,0). We can verify the CRC value for each possible frame: if only one frame has a correct CRC, then this frame is the correct frame. If both frames have a correct CRC, which is possible but unlikely, then the frame cannot be decoded. Similarly, the possible frames for $n_1$ are either (3,4,1,5,0), (3,4,1,6,0), (3,4,1,5,6) or (3,4,1,6,6). Since there are more uncertainties, the probability of having at least two frames with a correct CRC is higher, and it is less likely that this frame can be decoded. In order to avoid having to compute a large number of CRCs (with limited decoding performances), we set a limit to how many CRCs are performed per frame.

In order to show the performance of using the CRC in our MAC protocol, we consider the following scenario. We force situations where Algorithm~\ref{algorithm:proposition2} occurs by ensuring that all end-devices send a colliding frame with a slight desynchronization. We set the frame size to 50 bytes, the SF to 7 and we set the number of CRC attempts per frame to 4 or 100. We implemented random symbols for the frames, and the actual CRC algorithm of the LoRaWAN standard, which is CCITT-16 (see Subsection~15.2 of~\cite{lorawan-2017}).

\begin{figure}[htbp]
	\centering
	\psfrag{x}[c]{Number of frames in collision}
	\psfrag{y}[c]{Average number of CRC attempts per frame}
	\includegraphics[scale=1.40]{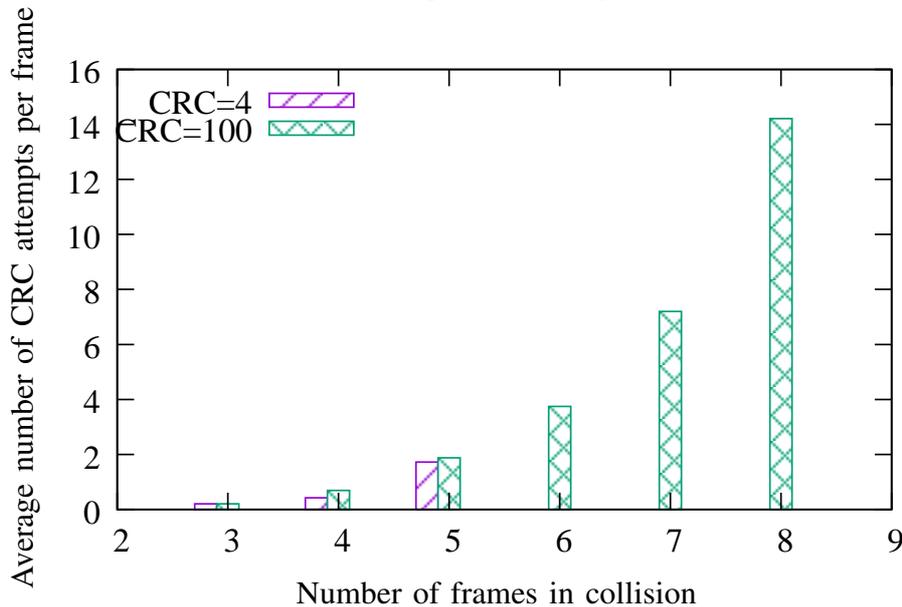}
	\caption{The number of CRCs needed to decode a frame increases with the number of frames in collisions.}
	\label{figure:crcRun}
\end{figure}

Figure~\ref{figure:crcRun} shows the average number of CRC attempts per frame. We notice that the number of needed CRC increases with the number of collided frames. This is because the more colliding signals, the more uncertainties there are in frames. Moreover, we notice that with a threshold of 4, the CRC algorithm cannot be applied for more than $n=5$ colliding frames, due to a large number of uncertainties. In this case, each frame needs more than 4 CRC to be decoded, so no CRC is actually computed. However, frames are able to be decoded by increasing the number of authorized CRCs per frame, e.g. to 100.

\begin{figure}[htbp]
	\centering
	\psfrag{x}[c]{Number of frames in collision}
	\psfrag{y}[c]{Number of decoded frames with CRC}
	\includegraphics[scale=1.40]{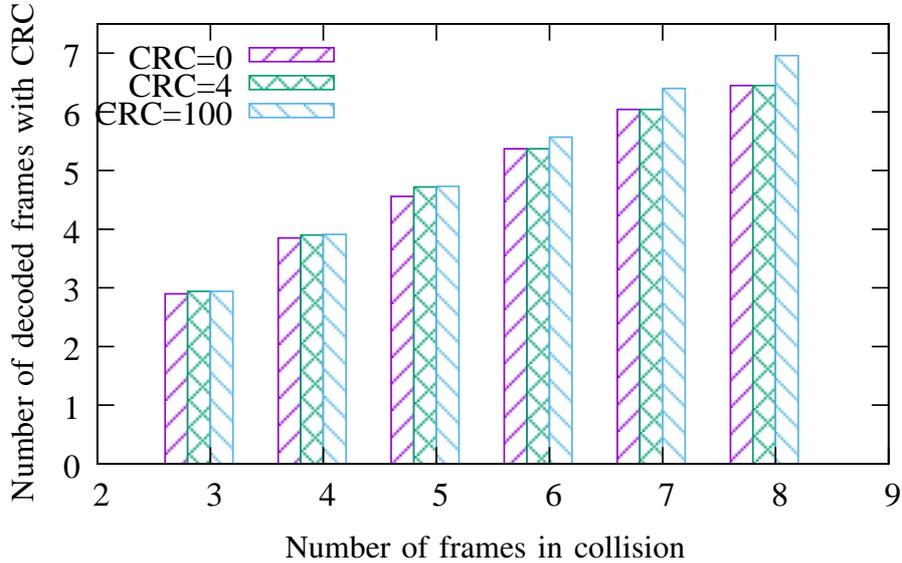}
	\caption{The CRC algorithm increases the number of decoded frames.}
	\label{figure:crcDecoding}
\end{figure}

Figure~\ref{figure:crcDecoding} shows the number of decoded frames with and without CRC for the scenario described above. We can notice that for a small number of authorized CRCs per frame such as 4, we see a small improvement when the number of colliding frames is less than or equal to 5. Above this number, the CRC algorithm is not able to decode frames and thus, it has the same behaviour as if CRC were disabled (case of CRC=0). However, by increasing the number of authorized CRCs per frame, we can notice that the number of decoded frames increases slightly and can improve the throughput up to 8\% when eight frames are colliding.

\section{Proposed Collision Resolving MAC protocol}
\label{section:new-mac}

In this section, we present a new MAC protocol which enables slightly desynchronized LoRa signals. Then, we provide an analysis of this proposed MAC protocol.

\subsection{Protocol Description}

Algorithm~\ref{algorithm:proposition} and Algorithm~\ref{algorithm:proposition2} require transmissions to be slightly desynchronized, by less than one symbol, which is a rare event in LoRaWAN. Thus, we designed a new MAC protocol called Collision Resolving-MAC (CR-MAC).

The CR-MAC protocol works as follows. Each gateway sends periodic beacons on each SF. These beacons are sent simultaneously by all gateways, as in Class B of LoRaWAN. Upon receiving a beacon, each end-device starts $S$ consecutive slots, whose duration is equal to the maximum frame transmission plus one symbol. To transmit a frame, an end-device has to wait for the beginning of a slot. It then draws a random number between 0 and $s=(SD/\delta)-1$, and delays its transmission by $s\times\delta$. We call sub-slots the possible starting times within each slot.

\begin{figure}[htbp]
    \centering
    \includegraphics[scale=1]{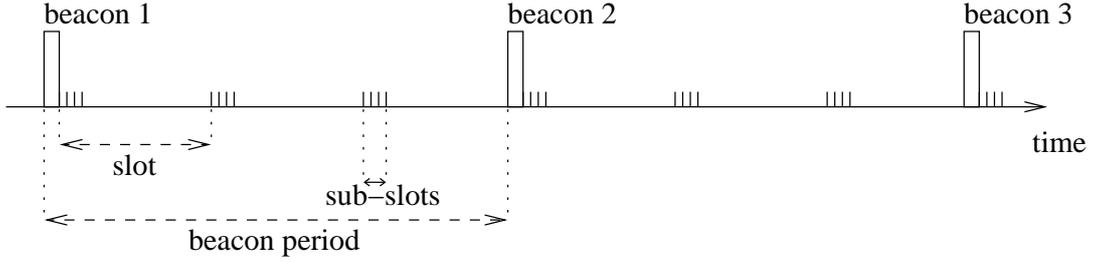}
    \caption{Our proposed CR-MAC protocol.}
    \label{figure:mac}
\end{figure}

Figure~\ref{figure:mac} depicts the CR-MAC protocol. There are three beacons, and $S=3$ slots after each beacon. At the beginning of each slot, there are $s=4$ sub-slots, which correspond to possible starting times for the transmission of frames.

With the CR-MAC protocol, if $n$ end-devices decide to transmit a frame on the same channel, with the same $SF$ and on the same slot, the probability that these transmissions are slightly desynchronized is equal to the probability that each node chooses a different sub-slot. This probability increases with $s$, and decreases with $n$.

If several end-devices transmit during the same sub-slot, Algorithm~\ref{algorithm:proposition2} fails. However, this can be detected by counting the number of frequencies in the set $newF$ in Algorithm~\ref{algorithm:proposition}: if it is equal to two or more, there are multiple transmissions in the same sub-slot.

In practice, the number of slots $S$ depends on the clock drift of the end-devices, and on the symbol duration. $S$ has an impact on the energy efficiency of CR-MAC, as it requires end-devices to listen to the beacon. Note that it would be possible for the node to listen to the beacon only when it has a frame to transmit, but in this case, $S$ would have a larger impact on the delay.

The number of sub-slots $s$ is computed as the symbol duration divided by $\delta$. The value of $s$ gives an upper bound on the number of frames that can be decoded by Algorithm~\ref{algorithm:proposition2}, or equivalently on the number of slightly desynchronized transmissions. Recall that $\delta$ is fixed by the hardware capabilities.

The impact of this protocol on the energy consumption is limited to end-devices listening to periodic beacons, and to a slight increase in the delay before a transmission (this delay is smaller than the slot duration).

\subsection{Analysis of the Proposed CR-MAC protocol}
\label{subsection:mac-analysis}

The performance of the CR-MAC protocol is closely related to the number of collisions, namely the number of end-devices choosing the same slot but different sub-slots.

If we denote by $n$ the number of end-devices that choose the same slot and by $s$ the number of sub-slots in the same slot, the probability that all $n$ end-devices choose a distinct sub-slot $p(n,s)$ can be determined as follows:
\begin{itemize}
	\item{if $n>s$: at least two end-devices will choose the same sub-slot, therefore $p(n,s)=0$},
	\item{if $n\leq s$: the number of possible patterns where all $n$ end-devices choose distinct sub-slots is equal to the number of arrangements of $n$ among $s$, defined by the number of combinations of $n$ elements among $s$ with ordering of $n$, i.e. $A_s^n=C_s^n\times n!$, and the total number of patterns is equal to $s^n$, giving:
\begin{equation}
	p(n,s)=\frac{A_s^n}{s^n}=\frac{s!}{(s-n)!s^n}.
\end{equation}
}
\end{itemize}

This probability depends on $s$, which is limited by the hardware, and on $n$, which depends on the total number of end-devices and on their duty-cycle. Thus, if $s$ is small, it is important to reduce $n$ for CR-MAC to achieve a good performance.

Table~\ref{table:analysis} shows the probability that all end-devices have different sub-slots, for several values of $n$ and of $s$. Obviously, the probability is 0 when there are more end-devices than sub-slots. As the number of sub-slots increases, the probability increases. For instance, the probability that $n=4$ end-devices have different sub-slots is about 9\% for $s=4$, and is 41\% for $s=8$. As the number of end-devices increases, however, the probability decreases. For instance, for $s=8$, the probability decreases from 41\% for $n=4$ end-devices to about 2\% for $n=7$ end-devices. Thus, it is very important that the number of end-devices sharing the same slot is kept low, ideally between $2$ and $n=s/2$. In practice, this can be controlled by reducing the duty-cycle of end-devices.

\begin{table}[htbp]
    \centering
    \begin{tabular}{|c|c|c|c|c|c|c|}
        \hline
        Number of end-devices $n$ & Probability for $s=2$ & Probability for $s=4$ & Probability for $s=8$ \\
        \hline
        2 & 0.5 & 0.75 & 0.875\\
        3 & 0 & 0.375 & 0.656\\
        4 & 0 & 0.094 & 0.410\\
        5 & 0 & 0 & 0.205\\
        6 & 0 & 0 & 0.077\\
        7 & 0 & 0 & 0.019\\
        8 & 0 & 0 & 0.002\\
        \hline
    \end{tabular}
    \caption{Numerical application for the probability that the $n$ end-devices having chosen the same slot also choose different sub-slots, as a function of $n$ and of the number of sub-slots $s$.}
    \label{table:analysis}
\end{table}

\section{Numerical results}
\label{section:results}

In this section, we evaluate and compare the network performance in terms of system throughput, energy efficiency, and system delay for both the conventional LoRaWAN protocol and our CR-MAC protocol.

\subsection{Parameter settings}

Simulations are carried out using our own simulator developed in Perl. We considered only one network server and one gateway in the network. We set the number of allowed CRC computation per frame to 4 and the size of preamble for each frame to 6 symbols (which becomes 10.25 after the addition of 2 symbols for the sync word and 2.25 symbols for down-chirps). For some simulations, we vary the number of end-devices but we set the size of the sent frames to 50 bytes. For other simulations, we vary the size of sent frames but we set the number of end-devices to 100. We also consider that all end-devices have a duty cycle of 1\% and are on the same channel with the same SF without capture conditions\footnote{Note that, under capture conditions the performance of the network will be improved for both LoRaWAN and CR-MAC protocols.}. We set the bandwidth to 125~kHz in order to have a fair comparison of the delay as it depends on the bandwidth and on the SF~\cite{lora}. We also set the number of slots in a beacon period to 100 for our CR-MAC protocol unless otherwise specified and we consider a beacon size of 10 bytes. Finally, in case of collision, we consider only one retransmission for successful reception, which is an ideal condition for conventional LoRaWAN\footnote{Note that a successful decoding after the first retransmission is more likely to happen with CR-MAC than with LoRaWAN, as shown later in Subsection~\ref{subsection:throughput}.}.

\subsection{Throughput}
\label{subsection:throughput}

\begin{figure}[htbp]
	\centering
	\psfrag{x}[c]{Size of frames (in bytes)}
	\psfrag{y}[c]{\% successful decoding}
	\includegraphics[scale=1.40]{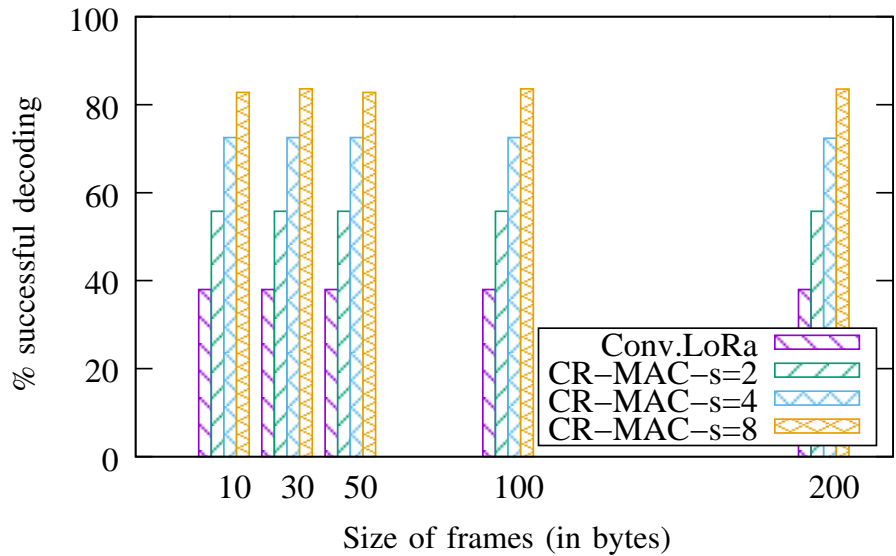}
	\caption{The percentage of successfully decoded frames increases by increasing the number of sub-slots.}
	\label{figure:throughputVsFrameLength}
\end{figure}

Figure~\ref{figure:throughputVsFrameLength} shows the percentage of successfully decoded frames as a function of the size of the sent frames for both the conventional LoRaWAN and our MAC protocol called CR-MAC. We vary the number of sub-slots to 2, 4, and 8. In the conventional LoRaWAN protocol, when several transmissions overlap, the signals collide and are thus considered lost. However, in the CR-MAC protocol, when more than one end-device use the same slot, their signals might be decoded if each end-device uses a different sub-slot. Thus, increasing the number of sub-slots reduces destructive collisions and increases the throughput of the system. It can be seen that the CR-MAC protocol outperforms the conventional LoRaWAN protocol. Indeed, the throughput computed by LoRaWAN is about 40\%, while it reaches 58\% using CR-MAC with two sub-slots, 76\% with four sub-slots, and 83\% with eight sub-slots.

\begin{figure}[htbp]
	\centering
	\psfrag{x}[c]{Number of end-devices}
	\psfrag{y}[c]{\% successful decoding}
	\includegraphics[scale=1.40]{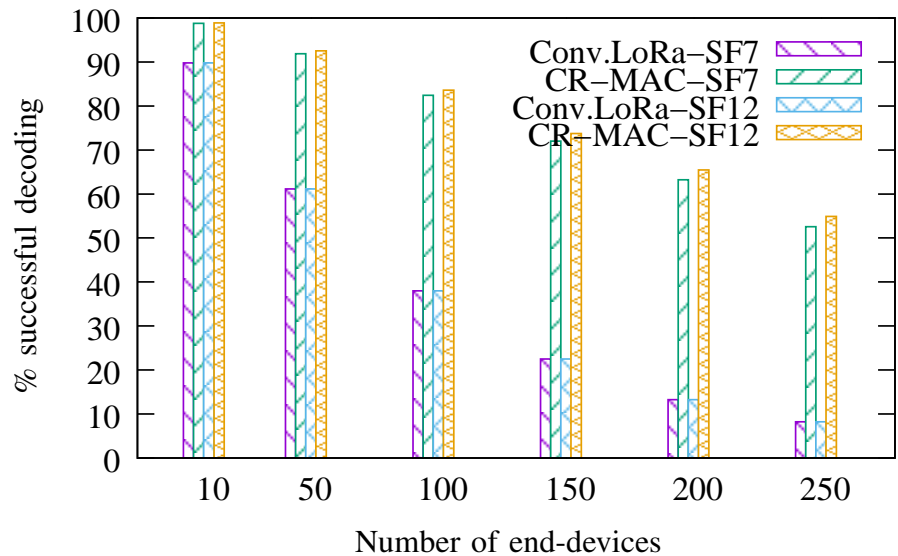}
	\caption{The collision resolution technique yields to considerably increasing the percentage of decoded frames.}
	\label{figure:throughputVsNodes}
\end{figure}

Figure~\ref{figure:throughputVsNodes} shows the percentage of successfully decoded frames in terms of the number of end-devices in the network for both LoRaWAN and CR-MAC protocols. We notice that this percentage decreases by increasing the number of end-devices for both protocols. This is due to the fact that in large networks, collisions are more important than in small networks. We can also notice that the performance of LoRaWAN degrades consistently compared to CR-MAC for both spreading factors SF7 and SF12. The percentage of loss is about 9 times lower in large networks (250 end-devices) than in small networks (10 end-devices). This percentage of loss is less drastic using CR-MAC. Indeed, in small networks, it is almost 0\% and even for large networks, the throughput of the system goes up to 52\% with SF7 and 55\% for SF12. This is due to the fact that, with our proposed superposed signal decoding technique, CR-MAC is able to resolve many colliding frames. 

\begin{figure}[htbp]
	\centering
	\psfrag{y}[c]{Average throughput (bps)}
	\psfrag{x}[c]{Number of end-devices}
	\includegraphics[scale=1.40]{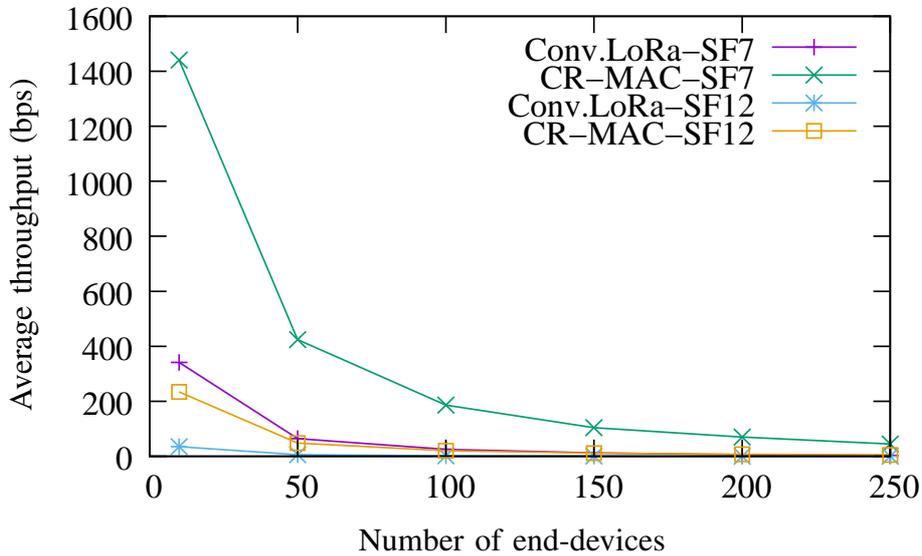}
	\caption{The collision resolution technique yields to considerably increasing the average throughput of the network.}
	\label{figure:debit}
\end{figure}

Figure~\ref{figure:debit} shows the average throughput in terms of the number of end-devices in the network for both LoRaWAN and CR-MAC protocols. We notice that the average throughput decreases with the number of end-devices, as in large networks, collisions occur more frequently. Moreover, we notice a large gain between the throughput computed with SF7 and the throughput computed with SF12. Indeed, although a larger SF enables slightly better collision resolution, it corresponds to much smaller bitrates. Finally, as our protocol is able to decode superposed LoRa signals, we notice that it outperforms LoRaWAN protocol with a gain up to 75\%.

\subsection{Energy Efficiency}
\label{subsection:energy-efficiency}

\begin{table}
	\centering
	\begin{tabular}{|c||c|c||c|c||}
	\hline
	Nb. End-devices & Conv.LoRa SF7 & CR-MAC SF7 & Conv.LoRa SF12 & CR-MAC SF12\\ 
	\hline
	10 & $5428.571\times10^3$ & $11000\times10^3$ & $28.16\times 10^3$ & $173.945\times10^3$ \\
	50 & $203.301\times10^3$ & $1107.754\times10^3$ & $1049$ & $3290$ \\
	100 & $40.761\times10^3$ & $267.908\times10^3$ & $209$ & $1655$ \\
	150 & $12.772\times10^3$ & $103.184\times10^3$ & $65$ & $619$ \\
	200 & $5.055\times10^3$ & $52.921\times10^3$ & $25$ & $316$ \\
	250 & $2.388\times10^3$ & $27.777\times10^3$ & $12$ & $163$ \\
	\hline
	\end{tabular}
	\caption{The energy efficiency (in bpJ) is reduced by increasing the number of end-devices in the network.}
	\label{table:eevsnodes}
\end{table}

In this subsection, we compute the energy efficiency defined by the ratio of the total number of successfully received bits and the total consumed energy. The consumed energy for LoRaWAN is the sum of transmit powers during frame transmission for all the end-devices. However, for CR-MAC, the consumed energy is the sum of the transmit powers during frame transmission for all end-devices and the power required for listening beacons during the time on air of beacons for a beacon size of 10 bytes.

Table~\ref{table:eevsnodes} shows the energy efficiency computed for both LoRaWAN and CR-MAC protocols, for SF7 and SF12. We set the number of slots to 100 and we vary the number of end-devices in the network. The transmit power is set to 66 mW and the received power is set to 19.5 mW~\cite{sx1272}. We notice that using SF7, CR-MAC protocol outperforms LoRaWAN with a gain up to 50\% for small networks and up to 90\% for large networks. Moreover, using SF12, the CR-MAC protocol shows a large gain compared to LoRaWAN. The gain goes up to 84\% for small networks and up to 92\% for large networks. This is due to the fact that the collision resolution implemented by CR-MAC reduces the delay as the number of retransmissions decreases compared to LoRaWAN, and considerably increases the average throughput.

\begin{figure}[htbp]
	\centering
	\psfrag{EE}[c]{Energy efficiency (kbpJ)}
	\psfrag{slots}[c]{Number of slots in a beacon period}
	\includegraphics[scale=1.40]{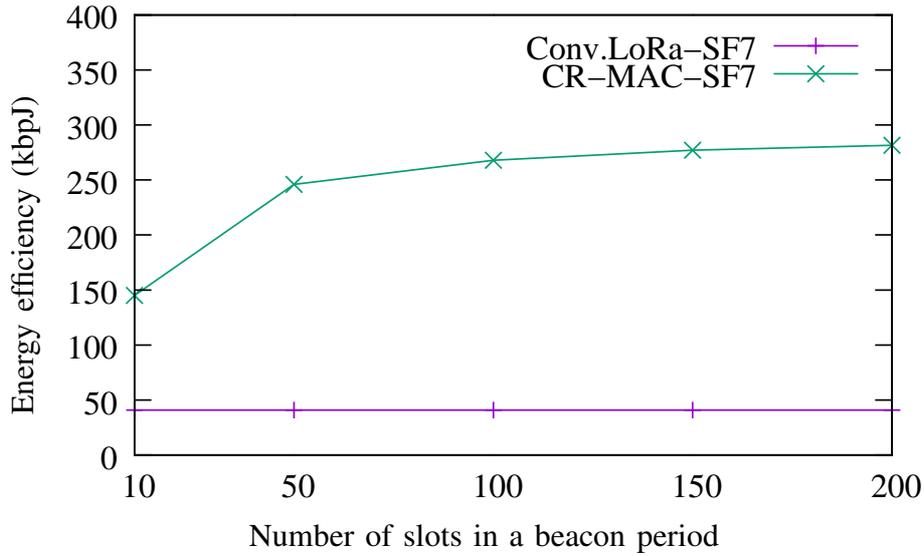}
	\caption{Using SF7, the transmission time is small, thus the end-devices listen frequently. This reduces the energy efficiency. Despite this, CR-MAC outperforms LoRaWAN.}
	\label{figure:eevsslotssf7}
\end{figure}

\begin{figure}[htbp]
	\centering
	\psfrag{EE}[c]{Energy efficiency (kbpJ)}
	\psfrag{slots}[c]{Number of slots in a beacon period}
	\includegraphics[scale=1.40]{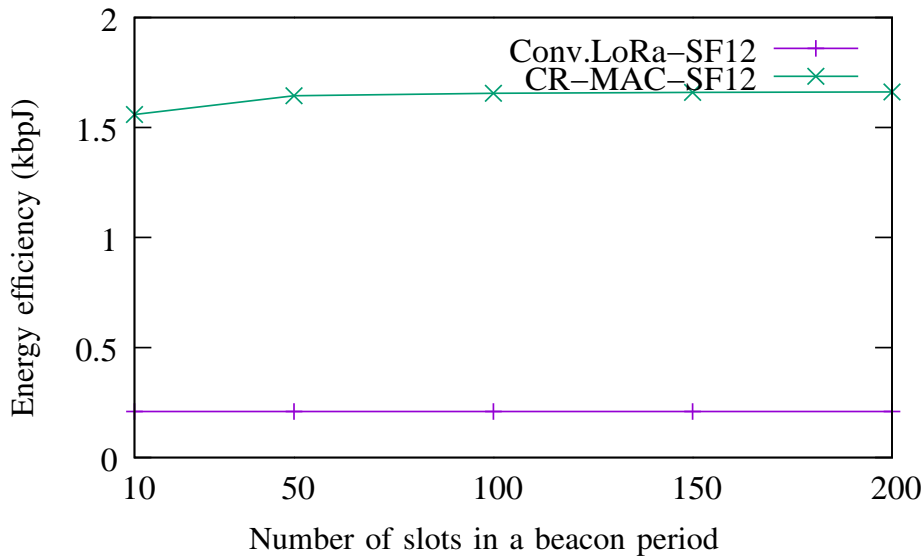}
	\caption{Using SF12, the transmission time is large, thus the listening period of the end-devices is short. This does not significantly impact the energy efficiency. CR-MAC outperforms LoRaWAN.}
	\label{figure:eevsslotssf12}
\end{figure}

Figure~\ref{figure:eevsslotssf7} and Figure~\ref{figure:eevsslotssf12} show the energy efficiency computed for both LoRaWAN and CR-MAC protocols. Here, we set the number of end-devices to 100 and we vary the number of slots in the beacon period. The energy efficiency computed by LoRaWAN is always the same. This is because transmissions in LoRaWAN follow the ALOHA mechanism and are independent of slots. However, we notice that the energy efficiency computed by CR-MAC slightly increases with the number of slots especially with SF7. Indeed, frames transmitted with SF7 have a short time on air. Thus, end-devices are in listening mode frequently and the CR-MAC protocol generates more beacon periods. This is why the energy efficiency is less important with a small number of slots than the energy efficiency achieved with a large number of slots. However, frames transmitted with SF12 have a large time on air and thus CR-MAC protocol results into less beacon periods per unit time, compared to SF7. This is why the energy efficiency remains almost constant against the number of slots in a beacon period. Compared to LoRaWAN, we observe a gain between 72\% for a small number of slots and 86\% for a large number of slots using SF7, and a gain of 87\% using SF12.

\subsection{Delay}
\label{subsection:delay}

\begin{figure}[htbp]
	\centering
	\psfrag{x}[c]{Number of end-devices}
	\psfrag{y}[c]{Delay (in seconds)}
	\includegraphics[scale=1.40]{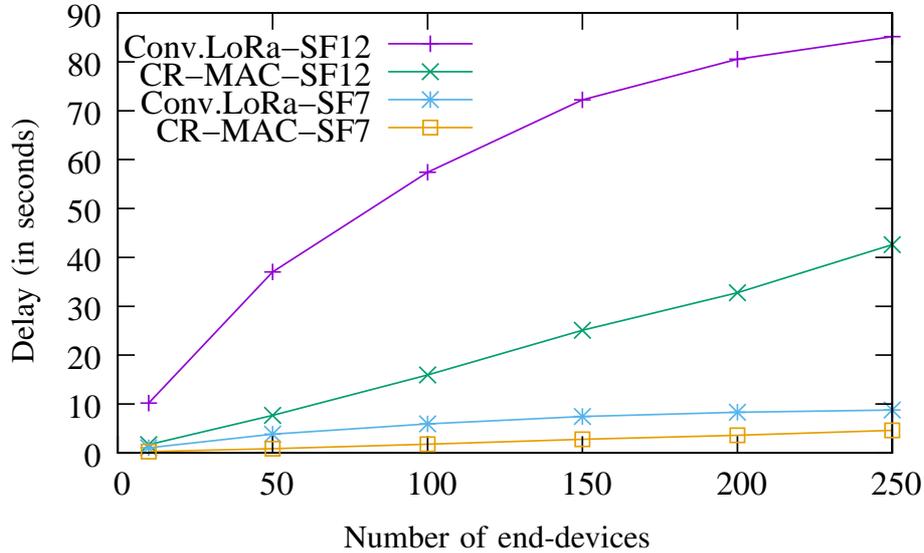}
	\caption{The delay against the number of end-devices for CR-MAC outperforms the delay for LoRaWAN due to our collision resolution technique.}
	\label{figure:delay}
\end{figure}

Figure~\ref{figure:delay} shows the average delay in terms of the number of end-devices for LoRaWAN and CR-MAC protocols. We notice that the delay increases with the number of end-devices and SF for both protocols. Indeed, in conventional LoRaWAN with a large number of end-devices, the probability to send frames without interference is low as end-devices use ALOHA mechanism for transmission. We observe also that CR-MAC outperforms LoRaWAN protocol and shows a delay reduction of 10\% for small networks and of up to 45\% for large networks. This is due to the fact that CR-MAC reduces the percentage of frame collisions as it is beacon-based. Moreover, CR-MAC is able to correctly decode collided frames which is not the case of LoRaWAN protocol, and thus CR-MAC may reduce the number of retransmissions. Furthermore, we notice a difference in delay when using different spreading factors. Indeed, as our protocol is able to cancel collisions, retransmissions are not always needed. In addition, the frame transmission duration depends on SF. With SF12, the transmission duration of a frame is larger than with SF7, thus inducing a larger delay for correct frames reception compared to SF7.

\begin{figure}[htbp]
	\centering
	\psfrag{x}[c]{Frame size (in bytes)}
	\psfrag{y}[c]{Delay (in seconds)}
	\includegraphics[scale=1.40]{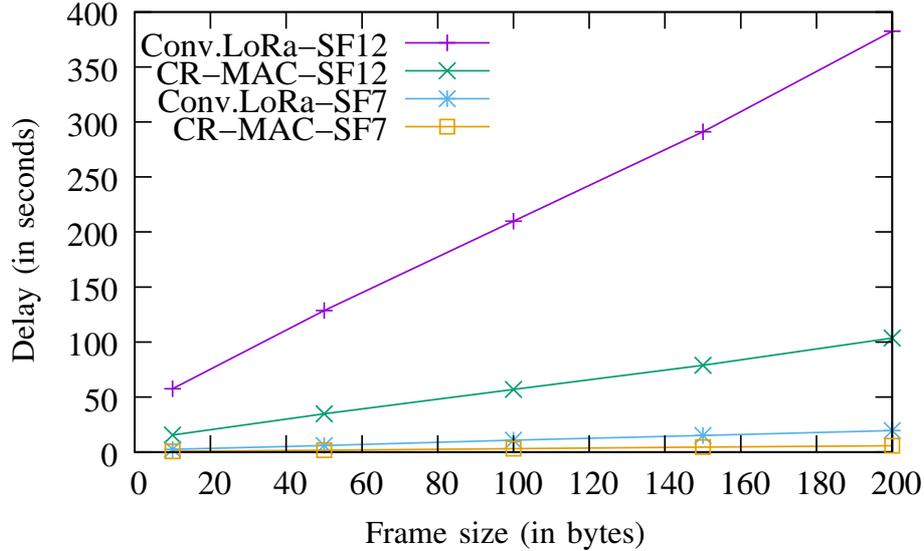}
	\caption{The delay against frame size for CR-MAC outperforms the delay for LoRaWAN due to our collision resolution technique.}
	\label{figure:delay-FrameLength}
\end{figure}

Finally, Figure~\ref{figure:delay-FrameLength} shows the average delay in terms of the size of the frames for LoRaWAN and CR-MAC protocols. We notice that the delay increases with SF and with the size of the frame for both protocols. Indeed, dealing with large frames yields to long transmissions and thus long duration for channel unavailability for each end-device. For example, for SF7, a frame of 10 bytes needs 39.17 ms to be transmitted, while a frame of 100 bytes needs 172.29 ms. Moreover, a frame of 50 bytes needs about 2 seconds to be transmitted with SF12, but it needs only 95 ms with SF7. The time on air of frames is the same for LoRaWAN and for CR-MAC, but CR-MAC requires an extra delay of half a slot (to wait for the slot start) plus half a symbol (to wait for the sub-slot start) on average. However, CR-MAC still outperforms LoRaWAN in Fig~\ref{figure:delay-FrameLength} even under ideal retransmission conditions. In reality, the gain may be even higher because LoRaWAN might use the 7 retransmissions defined in~\cite{lorawan-2015} which is not the case for CR-MAC as it is able to decode superposed signals using the collision resolution technique. We notice a reduction of the delay of 75\% for large frames.

To summarize, by resolving collisions, the CR-MAC protocol is able to jointly increase the network throughput and decrease the delay with a very small energy increment due to the beacons. Thus, the energy efficiency of CR-MAC is much higher than that of conventional LoRaWAN. In addition, smaller SFs further improve the achievable network performance.

\section{Conclusions}
\label{section:conclusion}

Collisions in LoRa networks are very harmful to the overall network performance. Indeed, when a gateway receives several superposed LoRa signals with comparable receive power levels, on the same channel and with the same SF, it is unable to decode these signals which are hence lost. In this paper, we proposed a novel beacon-based MAC protocol using a collision resolution technique that enables to decode two or more superposed LoRa signals. The proposed decoding algorithm exploits the slight desynchronization among superposed signals as well as the specificities of LoRa physical layer. We also show that the decoding performance of our collision resolution technique can be further improved by making use of the CRC which is already available in each frame. Simulation results show that, compared to the conventional LoRaWAN protocol, the proposed CR-MAC protocol provides remarkable performance improvements, both in terms of system throughput and energy efficiency. In addition, the proposed protocol enables significant delay reductions which is one of the most challenging tasks in 5G wireless communication systems. 

In the future work, we will further enhance our proposed protocol by designing tailored retransmission strategies. Furthermore, the feasibility of our proposal may be demonstrated through experimental evaluations using Software Defined Radio.

\newpage

\bibliographystyle{IEEEtran}
\bibliography{biblio}

\end{document}